\documentclass[graybox]{svmult}

\usepackage{type1cm}        % activate if the above 3 fonts are
\usepackage{makeidx}         % allows index generation
\usepackage{graphicx}        % standard LaTeX graphics tool
\usepackage{multicol}        % used for the two-column index
\usepackage{multirow}
\usepackage[percent]{overpic} %eigenes Package für Overpic
\usepackage[bottom]{footmisc}% places footnotes at page bottom
\usepackage{tikz} 
\usetikzlibrary{shapes.geometric, arrows}
\usepackage{pgfplots} 
\pgfplotsset{width=\linewidth, compat=1.6} 
\usepackage{pgfplotstable} 

\usepackage{newtxtext}       % 
\usepackage{newtxmath}       % selects Times Roman as basic font

\usepackage{booktabs}

\usepackage[normalem]{ulem} %only corrections (strikethrough)

\makeindex             % used for the subject index
\graphicspath{figures}

\usepackage{float}

\usepackage{subcaption}

\begin{document}

\title*{Direct Numerical Simulations of Droplet Impact onto Heated Surfaces using the Program Free Surface 3D (FS3D)}
% \author{Name of First Author and Name of Second Author}
\author{Manish Kumar, Rishav Saha, Johanna Potyka, Kathrin Schulte and Bernhard Weigand}
\authorrunning{M. Kumar, R. Saha, J. Potyka, K. Schulte and B. Weigand}
\institute{Manish Kumar and Rishav Saha \at Institute of Aerospace Thermodynamics (ITLR), University of Stuttgart, Pfaffenwaldring 31, 70569~Stuttgart, Germany, \email{manish.kumar@itlr.uni-stuttgart.de and rishav.saha@itlr.uni-stuttgart.de}} 
% \and Name of Second Author \at Name, Address of Institute \email{name@email.address}}
\maketitle
\abstract{
Droplet impact onto heated surfaces is a widespread process in industrial applications, particularly in the context of spray cooling techniques.
Therefore, it is essential to study the complex phenomenon of droplet spreading, heat removal from a hot surface, and flow distribution during the impact. 
This study focuses on Direct Numerical Simulation (DNS) of the initial stage of a water droplet impact onto a highly conducting heated surface, below the saturation temperature of the liquid. The maximum spreading diameters at different impact velocities in the presence of a heated surface, are analysed. Free Surface 3D (FS3D), an in-house code developed at the Institute of Aerospace Thermodynamics, University of Stuttgart, is used for this work. 
A grid independence study investigates the resolution required to resolve the flow field around the droplet. 
As evaporation effects during the initial stage of the droplet impact process are negligible, they are ignored. 
However, for longer simulation times, evaporation plays a significant role in the process. 
Preparing for such simulations, an evaporating droplet in cross flow is simulated to study the performance gain in the newly implemented hybrid OpenMP and MPI parallelisation and red-black optimization in the evaporation routines of FS3D.
Both the scaling limit and efficiency were improved by using the hybrid (MPI with OpenMP) parallelisation, while the red-black scheme optimization raised the efficiency only. 
%Therefore, OpenMP implementation and red-black optimization in the evaporation routine is undertaken, which increases the overall performance and scaling limit of the number of processing cores, for cases using a hybrid (MPI with OpenMP) parallelisation. 
%A comparative study is performed to measure the new version's performance compared to the original version. 
An improved performance of 23\% of the new version is achieved for a test case investigated with the tool MAQAO. 
Additionally, strong and weak scaling performance tests are conducted. The new version is found to scale up to 256 nodes compared to 128 nodes for the original version. The maximum time-cycles per hour (CPH) achieved with the new version is 35\% higher compared to the previous version.

}

\section{Introduction}
\label{sec:Introduction}
\vspace{-4mm}
The interaction between droplets and heated surfaces is a subject of significant interest across various scientific and technological fields. 
When a liquid droplet at room temperature contacts a hot surface, the temperature difference creates a heat flow from the surface toward the droplet. 
The heat flow gives a cooling effect on the surface, a beneficial effect commonly utilized in industrial applications such as spray cooling \cite{jia2003experimental}. 
The extent of heat transfer mostly depends on the temperature difference and the contact area. 
However, determining the contact area and the time scale associated with it is a well-known droplet dynamics problem. 
This process is influenced by numerous factors, including droplet parameters (such as diameter, impact velocity, and physical properties like saturation temperature, density, viscosity, and surface tension), surrounding gas parameters (such as pressure, temperature, and relative humidity), and surface characteristics (including wettability, diffusivity, surface roughness, and temperature) \cite{liang2017review, josserand2016drop}. 
Among these factors, the droplet impact velocity and the surface temperature are the most crucial parameters governing the impact dynamics and the heat transfer process \cite{bernardin1997mapping}. 
Hence, further investigations are required to understand the fundamentals of the droplet impact onto heated surfaces.

Under atmospheric conditions, a drop impacting perpendicularly to a flat wall will show a splashing, spreading, or bouncing behavior \cite{breitenbach2018drop}, depending on the impact velocity. 
Splashing occurs when a droplet impacts a surface at a high velocity, causing the droplet to break apart.
%Spreading occurs when a droplet impacts a surface at lower velocities. 
In case of lower impact velocities, however, the droplet spreads out smoothly over the surface without breaking apart into smaller ligaments. 
This results in a larger contact area between the droplet and the surface, enhancing the efficiency of the heat transfer \cite{bonn2009wetting}. 
Therefore, investigating the spreading phase is crucial.
Based on the surface temperatures, four distinct regimes can be identified: film evaporation, nucleate boiling, transition boiling and film boiling \cite{naber1993hydrodynamics, ko1996experiment}. 
%The heat transfer increases as the surface temperature increases, both in film evaporation and nucleate boiling regimes. 
%This reaches a maximum value during the transition from nucleate boiling to transition boiling regime, commonly known as the Critical Heat Flux (CHF) point. 
%Subsequently, as the surface temperature increases further, the heat transfer rate decreases due to the formation of vapor bubbles, restricting the heat transfer.
%Finally, the film boiling regime starts at the Leidenfrost temperature of the liquid, the point of minimum heat transfer \cite{naber1993hydrodynamics, ko1996experiment}. 
When the surface temperature is below the saturation temperature of the liquid (film evaporation), the heat transfer is dominated by heat conduction. Surface temperatures above this range lead to vapour bubbles and boiling. 
Therefore, to investigate the wetting behaviour fundamentally, it is worth considering the film evaporation regime, where boiling effects can be ignored.
%The evaporation time of a droplet is mostly influenced by the surface temperature. 

According to the droplet evaporation time, the film evaporation regime is commonly divided into three stages, initial, fundamental and final \cite{hu2002evaporation, rymkiewicz1993analysis}.
In the initial stage, the spreading, recoiling and oscillation of the droplet is observed, until it forms a stable spherical liquid cap. 
Subsequently, the three-phase contact line remains pinned, which involves cap evaporation with diminishing contact angle and droplet height.
Finally, the residual liquid rapidly shrinks and complete evaporation is observed. 
Studies show that the initial stage of droplet impact is faster and dominated by inertia \cite{rymkiewicz1993analysis}. 
During this stage, the droplet reaches its maximum contact area, significantly influencing the overall heat transfer rate. 
Therefore, the initial stage, where maximum wetting occurs, is considered critical for studying the process.
In addition, during this period, internal convection \cite{ruiz2002evaporation} of the droplet, airflow around the liquid \cite{berberovic2011inertia} also plays a significant role in the process.
%Many studies have also shown that uneven droplet temperatures during the impact process can result in internal convection and the Marangoni flow \cite{palmetshofer2024thermocapillary,pan2010symmetry,bhardwaj2010interfacial}.

In the present work, a single droplet of pure water at atmospheric conditions impacts perpendicularly onto a smooth isothermal surface at $363$~K. 
The study focuses on the initial stage of the impact process to capture the maximum spreading at high Weber and Reynolds numbers. %, internal convection \cite{ruiz2002evaporation}, air entrapment \cite{berberovic2011inertia} and airflow around the liquid.
These simulations are challenging due to the wide range of time and length scales involved. %, as well as the presence of high-velocity micro jets and heat conduction. %
This requires high spatial and temporal resolutions, which can only be achieved with an efficient parallelization of the code and the use of high-performance computing clusters. 
%Therefore, for a better understanding and prediction of the process, highly resolved Direct Numerical Simulations (DNS) are necessary. 
The DNS is conducted using the in-house multiphase flow solver FS3D running on a supercomputer of the High Performance Computing Center Stuttgart (HLRS), the HPE Apollo (Hawk). %Optimization and parallelization of evaporation routines are done to improve the efficiency of FS3D in cases where evaporation is significant to simulate (in future work) droplet wetting on heated surfaces.
Later in the report, Section 4  presents the computational performance gains achieved through newly implemented optimization and parallelization techniques in FS3D evaporation routines. Evaporation routines are required for future work in cases where evaporation cannot be ignored for droplet wetting on heated surfaces.

\vspace{-4mm}

\section{Methods}  \label{sec:methods}
\vspace{-2mm}

\subsection{Governing equations}  \label{sec:methods}
\vspace{-4mm}

The conservation equations solved for mass and momentum are given for this one-field approach by,
\begin{equation}
\rho_t + \nabla \cdot (\rho \mathbf{u}) = 0 ,
\end{equation}

\begin{equation}
[\rho \mathbf{u}]_t + \nabla \cdot [\rho \mathbf{u}\mathbf{u}] = \nabla \cdot [\mathbf{S} - \mathbf{I}p] + \rho\mathbf{g} + \mathbf{f}_\gamma ,
\end{equation}
with $\mathbf{u}$ denoting the velocity vector, $\rho$ the density, $p$ the static pressure, $\mathbf{S}$ the viscous stress tensor and $\mathbf{g}$ the gravitational acceleration. $\mathbf{f}_\gamma$ represents the surface forces in the vicinity of the interface \cite{brackbill1992continuum}. An incompressible Newtonian fluid is assumed. The energy equation solved in its temperature form reads as \vspace{-1mm}
\begin{equation}
\frac{\partial (\rho c_p T)}{\partial t} + \nabla \cdot (\rho c_p T \mathbf{u}) = \nabla \cdot (k \nabla T) + q,
\end{equation}
where $k$ is the heat conductivity, $T$ is the temperature, $c_p$ is the specific heat at constant pressure and $q$ is a source term. 
The Volume of Fluid (VOF) method by Hirt and Nichols \cite{hirt1981volume} is applied to distinguish between the different phases by introducing the volume fraction,~$f$, 

\vspace{-1mm}
\begin{equation}
	\label{eq:fdefinition}
	f(\textbf{x},t) = \left\{
	\begin{array}{ll}
		0 & \text{in the disperse phase},\\
		(0,1) & \text{for interfacial cells},\\
		1 & \text{in the continuous phase}
	\end{array}\right.
\end{equation}
that represents the amount of the liquid in each cell. %
A transport equation for $f$ needs to be solved to obtain the phase distribution over time, 

\vspace{-1mm}
\begin{equation}
	\label{eq:ftransport}
	\partial_t f + \nabla \cdot \left( f \textbf{u} \right) = 0.
\end{equation}

{To increase the accuracy of the corresponding $f$-fluxes, the Piecewise Linear Interface Calculation (PLIC) method by Rider and Kothe \cite{rider1998reconstructing}, which calculates the orientation and position of the interface in each cell is used.}

\subsection{Simulation tool: FS3D}  \label{sec:methods}
\vspace{-4mm}

The DNS is conducted using the in-house multiphase flow solver FS3D \cite{eisenschmidt2016direct} running on a supercomputer of the High-Performance Computing Center Stuttgart (HLRS), HPE Apollo Hawk platform. 
The incompressible Navier-Stokes equations and the energy equation \cite{eisenschmidt2016direct,Rieber2004,reutzsch2020} in temperature formulation are solved using the Finite-Volume method, where the interface is captured using the VOF method with the PLIC reconstruction \cite{hirt1981volume} in a Cartesian grid. 
Separate temperature fields are solved for both phases \cite{schlottke2010direkte}.
The continuum surface force (CSF) model \cite{brackbill1992continuum} is used to account for surface tension, while the interface reconstruction is performed using the piecewise linear interface calculation (PLIC) method \cite{rider1998reconstructing}. 
FS3D has been developed at the Institute of Aerospace and Thermodynamics for more than 25 years and has been applied successfully to various multiphase flow problems like droplet impacts onto structured surfaces \cite{ren2021air}, evaporating droplets \cite{schlottke2009vof}, thermocapillary flows \cite{ma2011direct} and drop-film interactions \cite{fest2021multiple}.
To simulate such processes, a high spatial and temporal resolution as well as large computational domains are necessary.
Therefore, the code is parallelized with MPI and OpenMP, and it is continuously optimized to increase its parallel efficiency and scaling on high-performance clusters. %

\vspace{-4mm}

\subsection{Computational Setup}
\label{sec:setup}
\vspace{-4mm}

\begin{figure}[ht]
	\begin{center}
		\begin{overpic}[width=0.45\columnwidth]{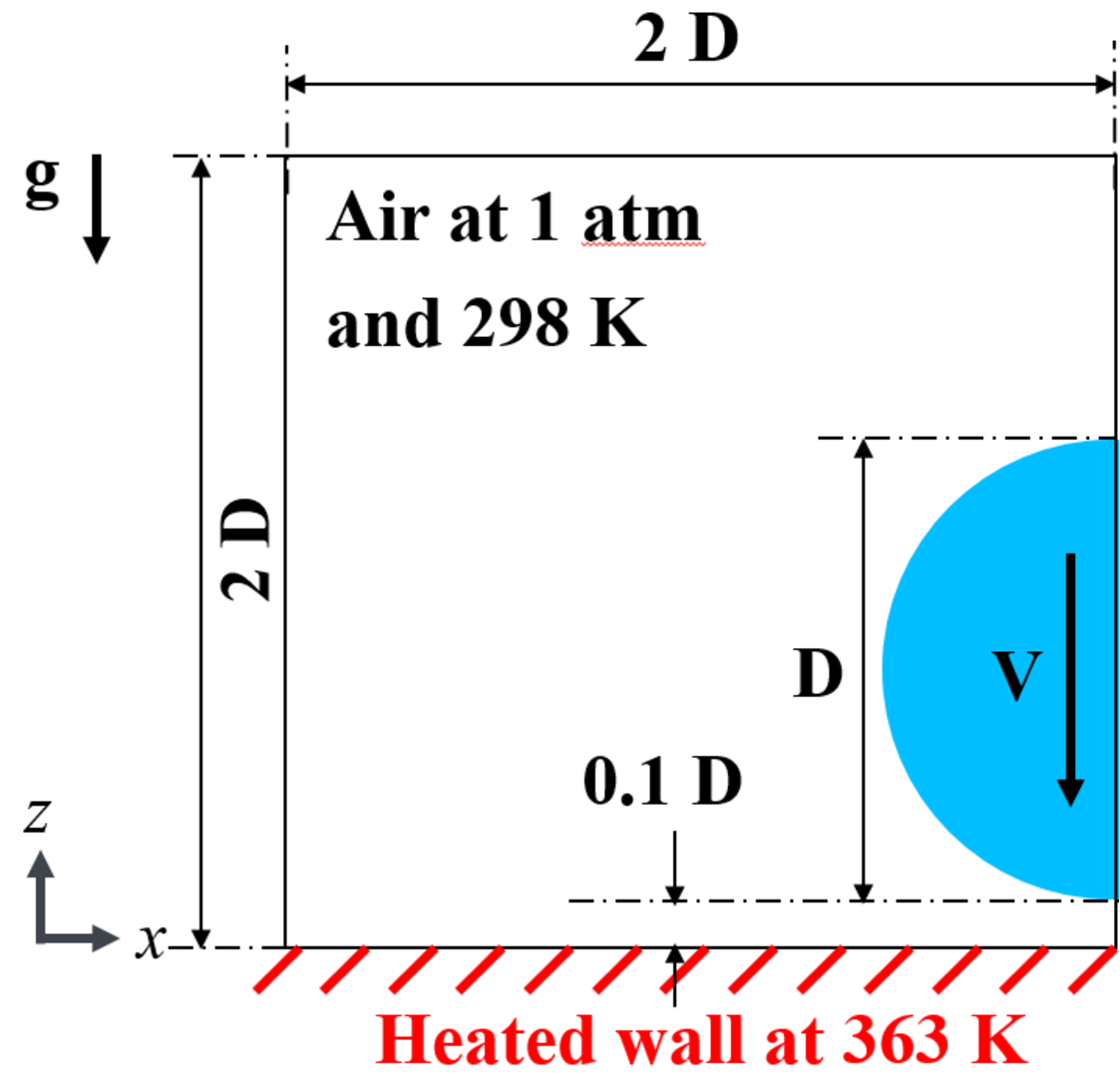}
		\end{overpic}
		\begin{overpic}[width=0.45\columnwidth]{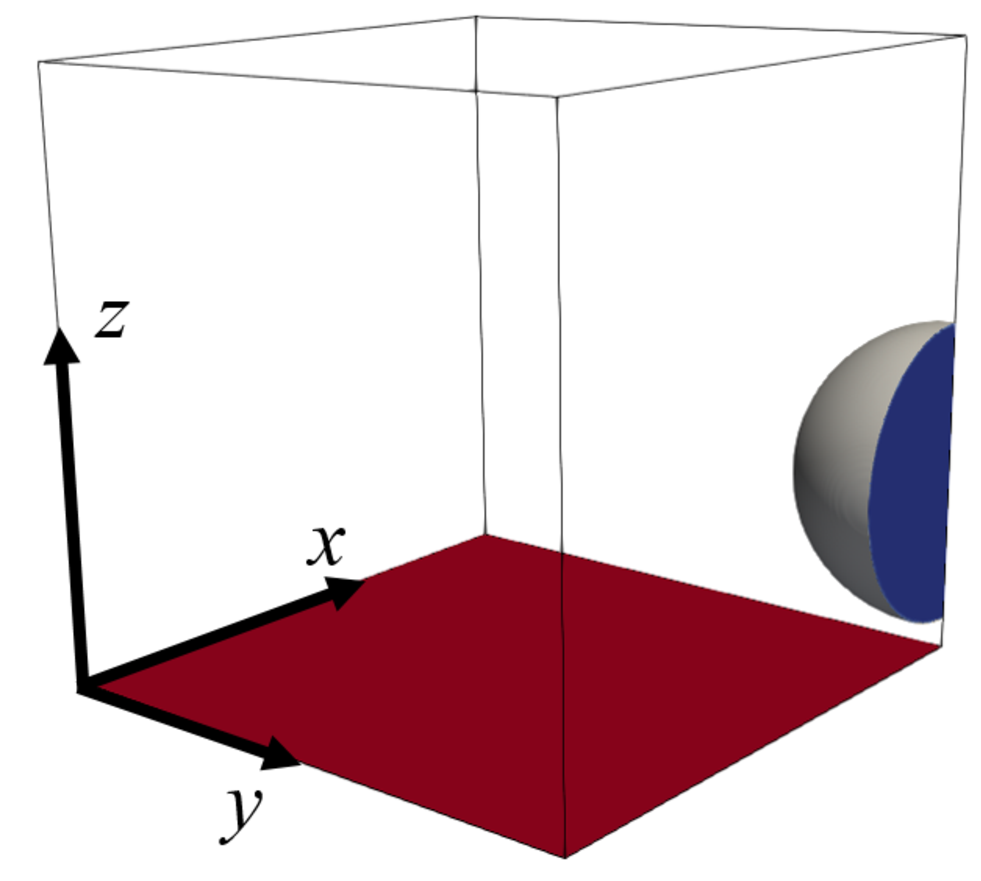}
		\end{overpic}
		%\vspace{1mm}
		\caption{(Left) Schematic 2D diagram of the computational domain with domain configuration at the initial point. (Right) 3D visualization of the computational domain showing a quarter of the droplet, with the hot surface highlighted in red.}
		\label{fig:Sketch}
	\end{center}

 \vspace{-6mm}
\end{figure}

A droplet of diameter $D$ is initialized in a cubic domain of size $2 D$ with an initial droplet velocity of $V$.
Initially, the droplet is at a distance of $0.1D$ above the bottom hot surface.  
The symmetry of the droplet impact dynamics is leveraged to study the behavior of droplets impacting a heated surface. 
Hence, a quarter droplet is simulated instead of the whole droplet.
Figure \ref{fig:Sketch} shows a two-dimensional schematic diagram and a three-dimensional view of the computational domain, detailing the initialized configuration.

\begin{table}[ht]
	\caption{Physical properties of air and water used, where $\rho$, $\mu$ and $c_p$  represent density, dynamic viscosity and specific heat at constant pressure, respectively. }
	\label{tab:physprop}
	\centering
	\begin{tabular}{p{1.8cm}p{2.5cm}p{2.5cm}p{2.5cm}}
		\hline\noalign{\smallskip}
		& $\rho \mathrm{(kg/m^3)}$   &  $\mu \mathrm{(Pa s)} $ & $c_p \mathrm{(kJ/(kg K)})$ \\
		\noalign{\smallskip}\svhline\noalign{\smallskip}
		Air & $1.184$ & $1.825\cdot 10^{-5}$ & 1.006 \\
		Water & $997$ & $8.921\cdot 10^{-4}$ & 4.18\\
		\noalign{\smallskip}\hline\noalign{\smallskip}
	\end{tabular}
	\vspace{-4mm}
\end{table}

A pure water droplet of spherical shape is considered to have an initial diameter of $D =2~\mathrm{mm}$. 
The numerical setup assumes that the fluid dynamics of water and surrounding air are incompressible and Newtonian. 
The properties of both fluids (air and water) are listed in Table \ref{tab:physprop}, and the surface tension coefficient ($\sigma$) is considered to be $72$~mN/m. 
At the moment, constant fluid properties are used because of the small temperature differences involved. The surrounding ambient of the droplet is at $1$ atmospheric pressure, and the initial temperature is $298$~K. 
The gravitational force ($g$) acts in the downward direction.
Continuous boundary conditions were applied to the far faces of the domain except for symmetric planes and the bottom surface.
The bottom surface is assumed to be of high conductive substrate, and an isothermal boundary condition is enforced at the temperature of 363 K, below the boiling temperature of the water. 
Evaporation is ignored as the time scale of evaporation is large compared to the wetting time scale. 
Droplet impact dynamics is studied at different impact velocities, and these values are listed in Table \ref{tab:case}, as well as the corresponding values of Weber number (We) and Reynolds number (Re).
We and Re are defined as We $=\rho_wV^2D/\sigma$ and Re $=\rho_wVD/\mu_w$, respectively, where the subscript w represents water.

\vspace{-2mm}

\begin{table}[ht]
	\caption{Investigated droplet impact velocities and corresponding values of Weber and Reynolds numbers.}
	\label{tab:case}
	\centering
	\begin{tabular}{p{1.8cm}p{1.8cm}p{1.8cm}p{1.8cm}p{1.8cm}p{1.8cm}}
		\hline\noalign{\smallskip}
		Cases & I  & II & III & IV & V\\
		\noalign{\smallskip}\svhline\noalign{\smallskip}
		V in $\mathrm{m/s}$ & $1$ & $1.5$ & 2 & $2.5$ & $3$ \\
		We & $27$ & $62$ & $110$ & $173$ & $250$\\
            Re & $2230$ & $3350$  & $4470$ & $5590$ & $6705$ \\
		\noalign{\smallskip}\hline\noalign{\smallskip}
	\end{tabular}
	\vspace{-8mm}
\end{table}

\vspace{-4mm}

\section{Results and Discussion}\label{sec:results}
\vspace{-2mm}

\subsection{Grid independence study}
\vspace{-4mm}

A grid independence study is presented in the following to determine the influence of the grid resolution on the simulations. 
Four grid sizes were investigated in the study. 
Details of the different grid sizes are listed in Table \ref{tab:grid}. 
Grid sizes were increased systematically by a factor of two, from 128 to 1024 in each direction of the cubic domain, which resulted in a systematic increase in the number of cells from two million to one billion. 
Hence, the number of cells per initial drop diameter was varied from 64 cells to 512 cells.

\begin{table}[ht]
	\caption{Details of different grids employed for the grid independence study.}
	\label{tab:grid}
	\centering
	\begin{tabular}{p{2.7cm}p{2.1cm}p{2.1cm}p{2.1cm}p{2.1cm}}
		\hline\noalign{\smallskip}
		   & GRID 128  & GRID 256 & GRID 512 & GRID 1024 \\
		\noalign{\smallskip}\svhline\noalign{\smallskip}
		  Grid size & 128$\times$128$\times$128 & 256$\times$256$\times$256 & 512$\times$512$\times$512 & 1024$\times$1024$\times$1024\\
		Number of cells & 2 million & 17 million & 134 million & 1 billion \\
            Resolution in $\text{cells}/D$ & 64 & 128  & 256 & 512 \\
		\noalign{\smallskip}\hline\noalign{\smallskip}
	\end{tabular}
	\vspace{-4mm}
\end{table}

\begin{figure}[ht]
	\begin{center}
		\begin{overpic}[width=0.49\columnwidth]{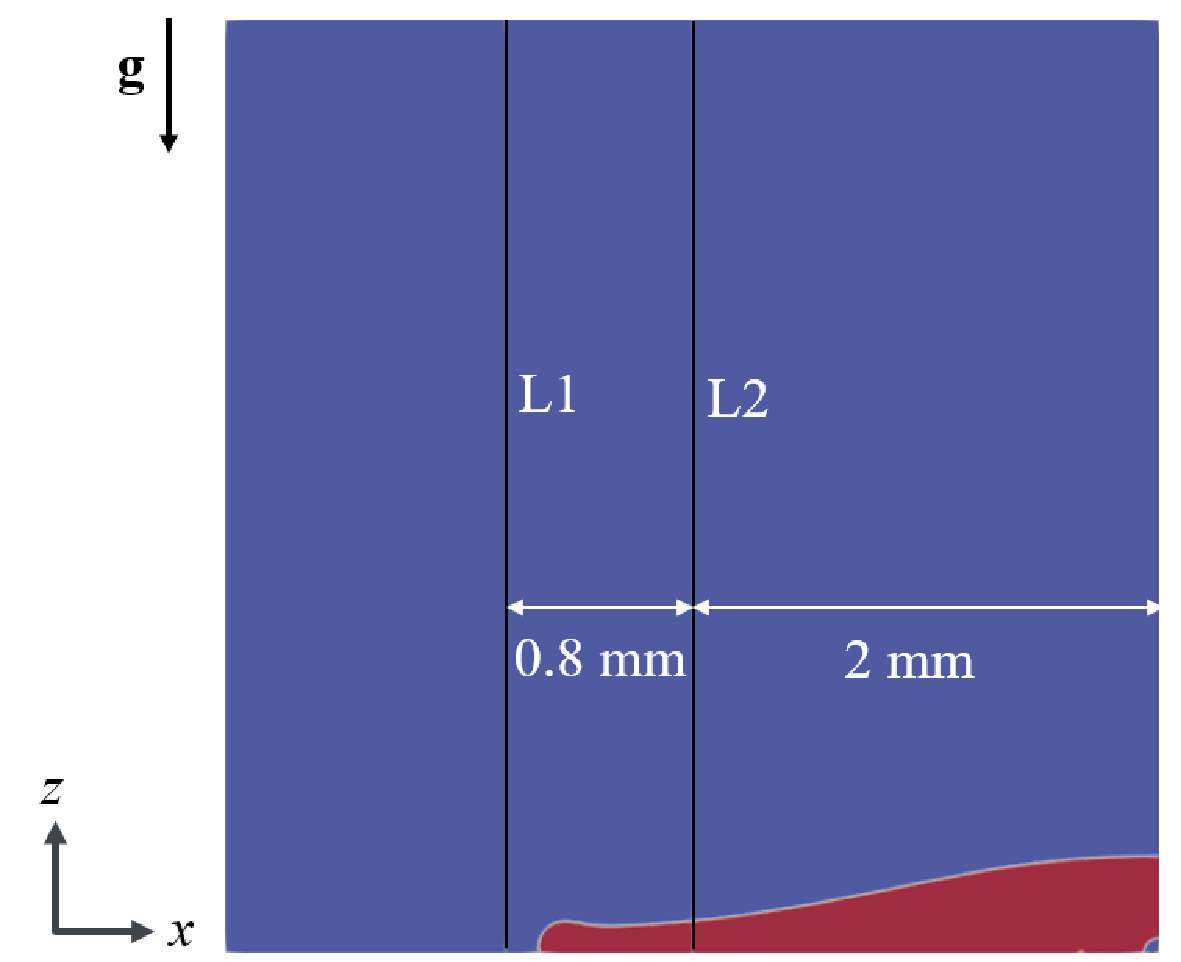}
		\end{overpic}
		\vspace{1mm}
		\caption{Snapshot of droplet simulation at time $t = 0.84~\mathrm{ms}$ indicating the location of probes L1 and L2 at which simulation parameters are compared for the grid independence study.}
		\label{fig:probe}
	\end{center}
 \vspace{-6mm}
\end{figure}

The case with the highest impact velocity is chosen to evaluate the grid independence. 
Figure \ref{fig:probe} shows a snapshot of one such simulation and indicates the locations of line probes L1 and L2. 
The location of probe L1 is placed outside the droplet (at $2.8~\mathrm{mm}$ from the right boundary), and probe L2 is placed inside the droplet (at 2 mm from the right boundary) so that both regions' velocity and temperature fields can be compared.
These fields are compared at an intermediate time, $t = 0.84~\mathrm{ms}$, when the droplet is still spreading.

\begin{figure}[ht]

\begin{subfigure}{0.49\textwidth}
\includegraphics[width=\linewidth]{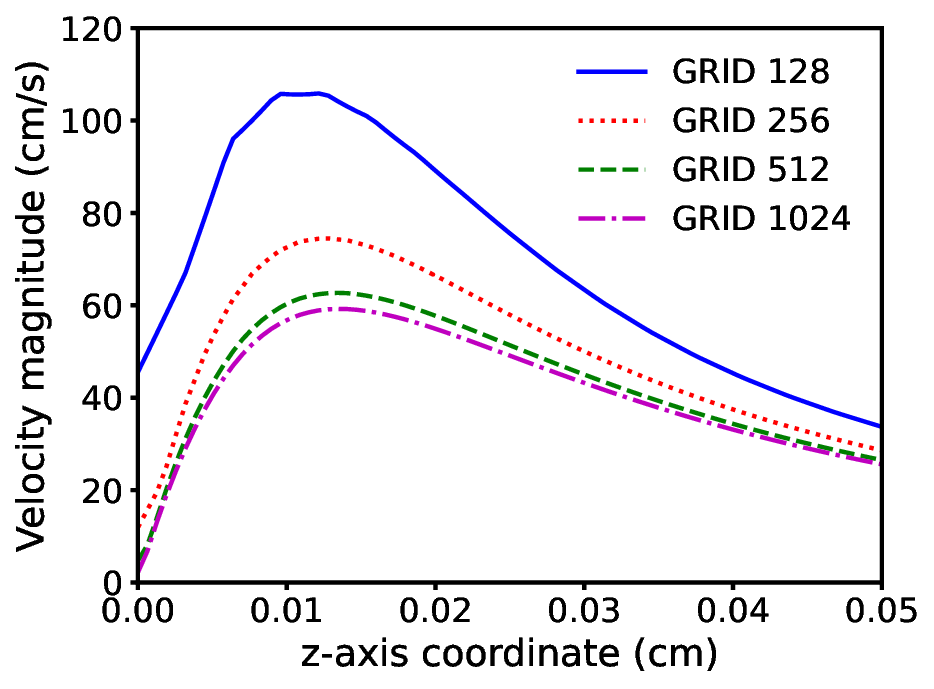}
\caption{}\label{gridv1}
\end{subfigure}
\hspace*{\fill}
\begin{subfigure}{0.49\textwidth}
\includegraphics[width=\linewidth]{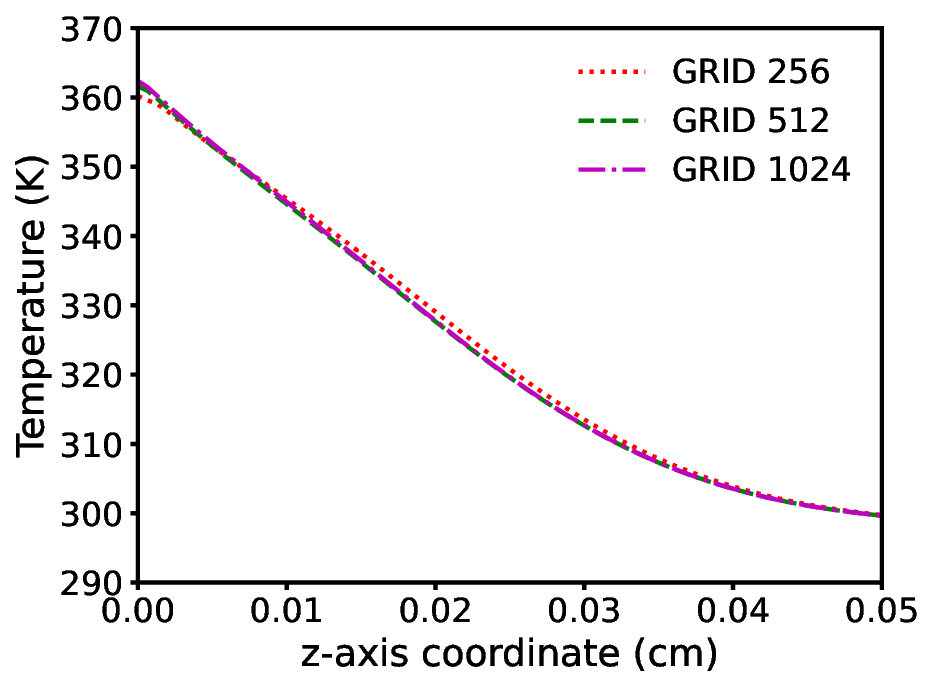}
\caption{}\label{gridt1}
\end{subfigure}

\caption{(a) Velocity magnitude and (b) temperature profile along z-axis calculated using different grid sizes at probe line L1.}
\vspace{-1mm}
\end{figure}

Figure \ref{gridv1} and \ref{gridt1} show the velocity magnitude and temperature variation along the z-axis and probe L1, in case of different grid sizes, respectively. 
For all grid sizes, the velocity magnitude increases, and starts to decrease after reaching a maximum value. 
However, temperature decreases monotonically and in a non-linear fashion. 
Also, it can be observed from Figure \ref{gridv1} that as the grid size increases from 128 to 1024, velocity profiles start to converge in the case of GRID 512 and GRID 1024. 
However, grid independence can be achieved from GRID 256 in the case of temperature, as shown in Figure \ref{gridt1}. Temperature data for GRID 128 is not plotted in figure \ref{gridt1}, as the temperature solution diverges for this coarse grid size. 

% \begin{figure}[ht]
% 	\begin{center}
% 		\begin{overpic}[width=0.48\columnwidth]{figures/grid_indp_V.png}
% 		\end{overpic}
% 		\begin{overpic}[width=0.48\columnwidth]{figures/grid_indp_T.png}
% 		\end{overpic}
% 		\vspace{1mm}
% 		\caption{(a) Schematic 2D diagram of problem setup with domain details, (b) 3D visualization of computational domain: a quarter droplet just before impacting on the heated surface.}
% 		\label{fig:Sketch}
% 	\end{center}
% \end{figure}

\begin{figure}[ht]

\begin{subfigure}{0.49\textwidth}
\includegraphics[width=\linewidth]{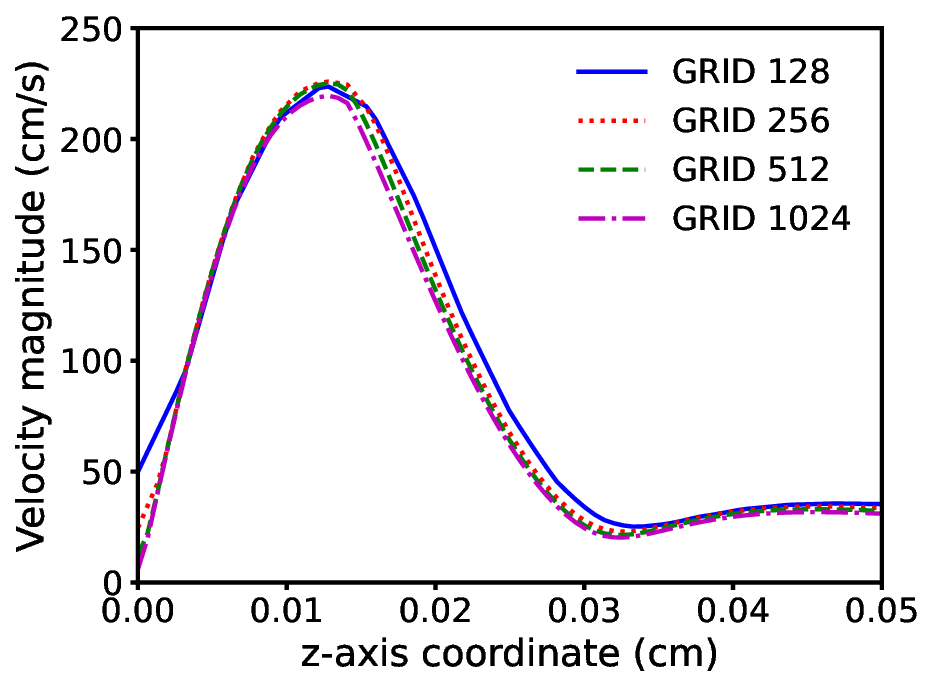}
\caption{}\label{gridv2}
\end{subfigure}
\hspace*{\fill}
\begin{subfigure}{0.49\textwidth}
\includegraphics[width=\linewidth]{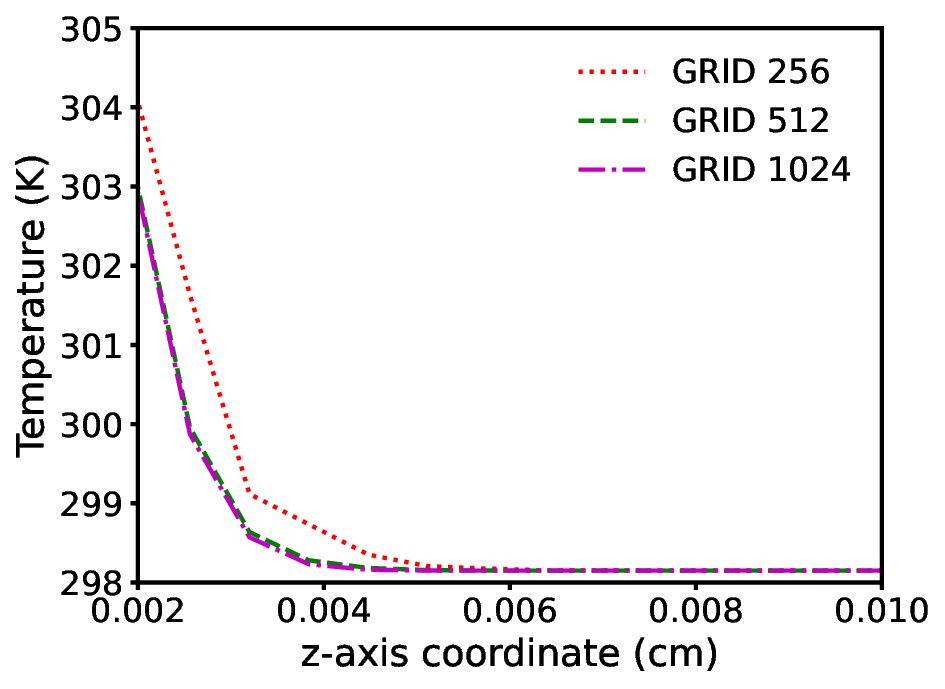}
\caption{}\label{gridt2}
\end{subfigure}

\caption{(a) Velocity magnitude and (b) temperature profile along z-axis calculated using different grid sizes at probe L2.}

\vspace{-4mm}
\end{figure}

Similarly, Figures \ref{gridv2} and \ref{gridt2} show the velocity magnitude and the temperature profile at probe line L2. The temperature in Figure \ref{gridt2} is plotted till 0.01 cm of z-axis coordinate within the droplet region where large differences can be seen and compared among different grid sizes.
Satisfactory grid independence is achieved from grid size GRID 256 in the case of velocity, as shown in Figure \ref{gridv2}. 
On comparing Figure \ref{gridv2} and \ref{gridv1}, it can also be observed that the velocity magnitude is higher inside the droplet compared to the ambient air. 
In the case of the temperature profile, grid independence was achieved from GRID 512. 
Hence, the present study uses GRID 512 for all simulations to study the droplet impact onto a heated surface.

\vspace{-4mm}

\begin{figure}[ht]
\centering
\includegraphics[width=0.99\linewidth]{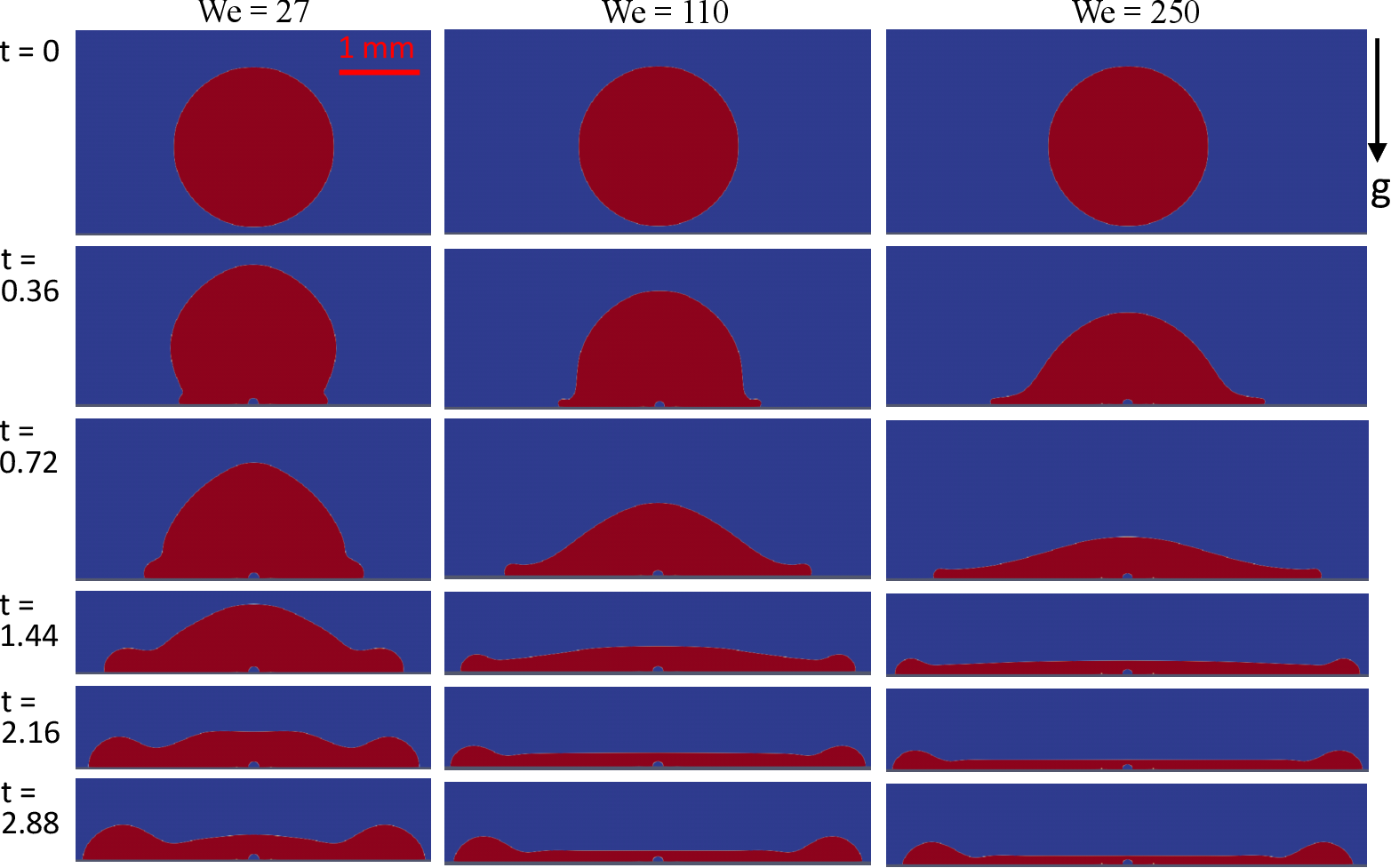}
\caption{Time-sequenced snapshots of droplet impact simulations for three cases of Weber numbers (i.e. 27, 110, and 250). The time (t) is provided in the first column of frames in milliseconds (ms).}
\label{dyna}
\vspace{-4mm}
\end{figure}

\subsection{Droplet spreading}
\vspace{-4mm}
Figure \ref{dyna} shows the wetting dynamics of droplets (in red color) impacting onto a heated surface in a quiet ambient air (in blue color). 
Time-sequenced images show how the droplets spread at different impact speeds, denoted by the corresponding Weber number. 
It can be observed from Figure \ref{dyna} that the droplet spreads rapidly until $0.72~\mathrm{ms}$, and after that, the spreading slows down until the droplet reaches its maximum spreading diameter.
Capillary waves can be observed at the liquid-gas interface of the droplet. 
As the droplet spreads, droplet liquid rushes towards the periphery of the droplet, creating a rim. 
The rim formation happens due to the high kinetic energy of the droplet impact, which tries to dominate over dissipating effects of droplet liquid, i.e., viscosity and surface energy. 
As the Weber number ($We$) increases, the droplet spreads faster and achieves a larger wetting diameter compared to an impact at a lower Weber number.

\begin{figure}[ht]

\begin{subfigure}{0.49\textwidth}
\includegraphics[width=\linewidth]{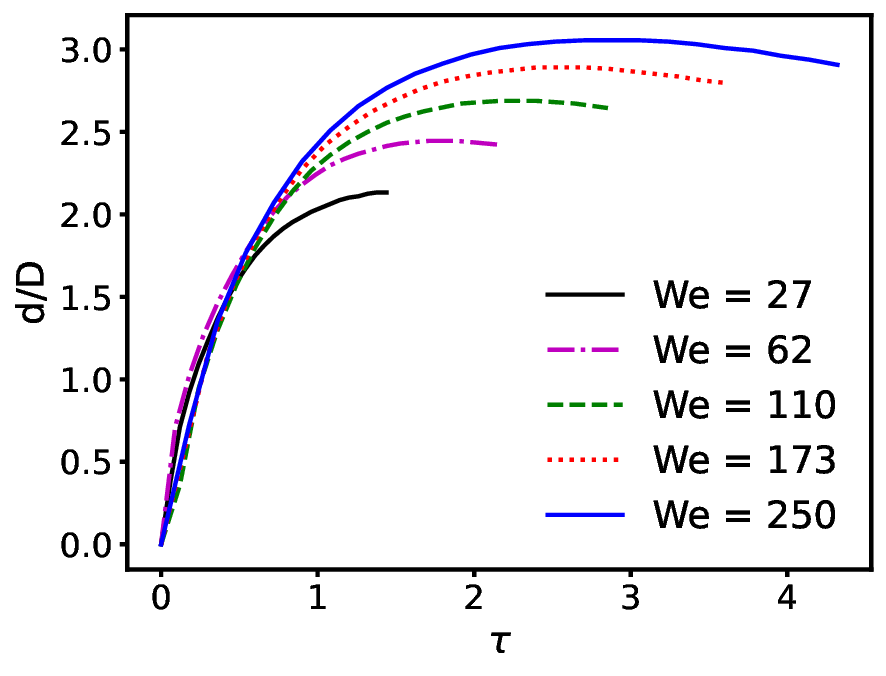}
\caption{}\label{dia_t}
\end{subfigure}
\hspace*{\fill}
\begin{subfigure}{0.49\textwidth}
\includegraphics[width=\linewidth]{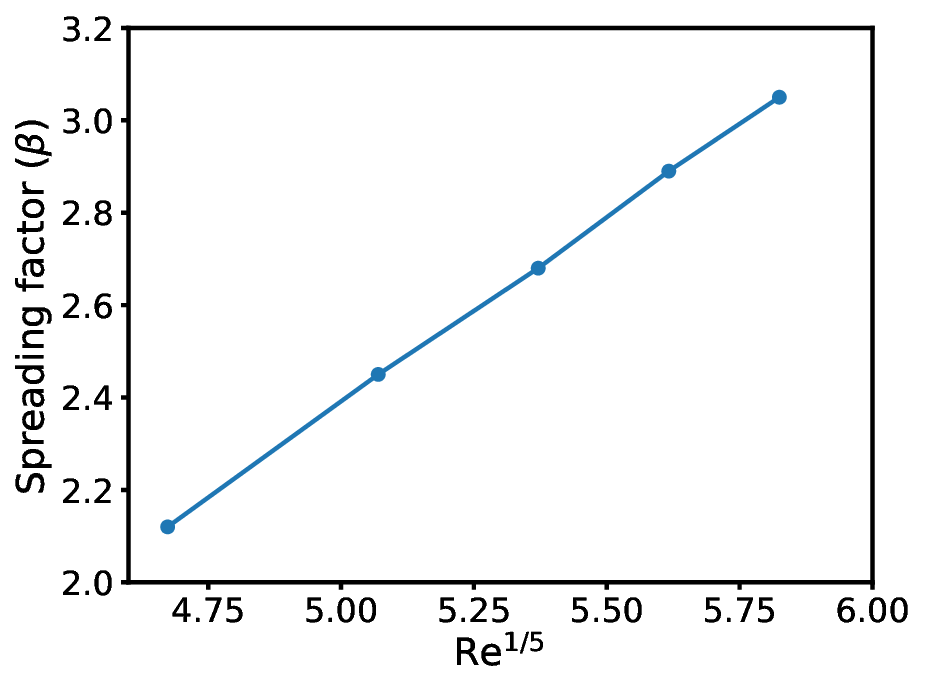}
\caption{}\label{reeffec}
\end{subfigure}

\caption{(a) Dimensionless wetting diameter (d/D) variation of the droplet with respect to dimensionless time ($\tau =$ t.V/D), for different cases of Weber numbers (We). (b) Variation of spreading factor ($\beta$) with respect to the Reynolds number (Re).}

\vspace{-4mm}
\end{figure}

The quantification of the dimensionless wetting diameter (d/D) variation with respect to dimensionless time ($\tau =$ t.V/D) for different Weber numbers (We) is shown in Figure \ref{dia_t}. 
The droplet wetting diameter increases non-linearly for all Weber numbers. 
As explained earlier, it can also be seen in Figure \ref{dia_t} that the maximum wetting diameter increases with increasing Weber numbers. 
In the case of a higher Weber number, the droplets attain the maximum wetting diameter earlier than in the lower Weber number cases. 
In the cases of $We = 173$ and $250$, the droplet attains the maximum wetting diameter around $1.5~\mathrm{ms}$ and starts to recede around $1.8$ to $2~\mathrm{ms}$. 
However, in the case of $We = 27$ and $62$, the droplet attains its maximum wetting diameter at $2.8~\mathrm{ms}$.

% \begin{figure}[ht]
% \centering
% \includegraphics[width=0.75\linewidth]{figures/diaVstime.png}
% \caption{Droplet dynamic wetting behavior: dimensionless wetting diameter (d/D) variation of the droplet with respect to dimensionless time ($\tau =$ t.V/D), for different cases of Weber numbers (We).}
% \label{dia_t}
% \end{figure}

% \begin{figure}[ht]
% \centering
% \includegraphics[width=0.75\linewidth]{figures/re_effect.png}
% \caption{Variation of spreading factor ($\beta$) of the impacting droplet with respect to the Reynolds number (Re).}
% \label{reeffec}
% \end{figure}

Figure \ref{reeffec} shows the effect of Reynolds number (Re) on the non-dimensional spreading parameter called the spreading factor ($\beta$). 
The spreading factor ($\beta$) is defined as the ratio of the maximum wetting diameter to the initial diameter of the spherical droplet before the impact. 
The variation of spreading factor ($\beta$) with respect to Re$^{1/5}$ is shown in Figure \ref{reeffec}. A linear dependence of $\beta$ can be observed, which indicates the viscous regime of droplet spreading \cite{liang2017review}.

From the heat transfer perspective, it is clear from figures \ref{gridt1} and \ref{gridt2}, that the maximum water temperature reached is significantly lower than the maximum air temperature. 
This is because the specific heat of water is greater than that of air. Additionally, only a thin portion of the liquid film, around 20\% of its height, is heated. 
This indicates that the heating process is slower than the droplet spreading.
In the future, more focus will be kept on the heat transfer part and the consideration of a possible influence of the temperature-dependent fluid properties.

All simulations with GRID 512 utilizes 8 MPI processes in each {\it{x}}, {\it{y}} and {\it{z}}-direction, which corresponds to a total of 512 MPI processes. 
Four nodes were used with 128 MPI processes for a wall-time of 11.4, 13.6, 15.6, 16.5, and 17.2 hours for Cases I, II, III, IV and V, respectively. 

\vspace{-4mm}

\section{Computational Performance} \label{sec:Com_Perf}
\vspace{-4mm}

Continuous enhancements of FS3D's performance are necessary to keep pace with advancements in super-computing hardware and the demands of increasingly larger simulation domains and resolutions. %all while ensuring efficient resource utilization.
Several performance optimization techniques within FS3D have previously focused on the most time-consuming routines of the hydrodynamics solver, including the multigrid (MG) solver, viscosity solver, and momentum advection solvers \cite{HLRSBericht2020,stober2023dns, HLRSBericht2022}. 
%In the past, optimizations are focused on pure hydrodynamics, the core part of the solver. 
%FS3D's strength is that it also allows the simulation of phase change. 
However, for the study of heat and mass transfer \cite{Reutzsch2019, reutzsch2020}, optimizing the phase transfer (evaporation) routines is crucial to enhance the overall performance. 
This is especially important when the evaporation timescale becomes comparable to other physical phenomena, such as droplet wetting.
In the latest version of FS3D, evaporation routines are optimized with a Red-black scheme and OpenMP implementation. 
%The following subsection provides an overview of the implementation.
%Hence, the following subsections present an optimization of routines that facilitate the simulation of evaporation.
\vspace{-6mm}

\subsection{Red-black scheme and OpenMP implementation in evaporation routines}
\vspace{-4mm}

In a previous work on optimising the hydrodynamics solver of FS3D, Potyka et al. \cite{HLRSBericht2022} found that a significant speed-up for loops employing red-black schemes can be achieved. %
If the order of operations is organized such that the data required for the second, black iteration does not have to be loaded from main memory, a speed-up by up to a factor of two is possible. %
This can be achieved by slice-wise progress of the red and black iterations through the three-dimensional arrays, where the black iteration follows the red iteration with an offset of one slice. % of the three-dimensional data. %
%An investigation of the most time-consuming routines
An analysis in the phase change solver identified that a red-black scheme as one of the most time-consuming routines. %
Hence, the above-described cache-blocking approach was adapted for the red-black Gauss-Seidel smoother solving for the conducted heat. %
Which resulted in a reduction of the computational time %for solving the heat conduction 
by 29\% for the test case described in Sec.~\ref{sec:test}. %
The obtained gain is below the theoretical value of $\approx50\%$ reduction in computational time achievable with perfect cache reuse. %
The equation solved in this routine requires eleven three-dimensional arrays, and seven of them are accessed with a seven-point stencil. %
Therefore, the benefit is lower than the optimum. %
Nevertheless, this re-organization of only one loop translates to a reduction of the overall compute time of FS3D by 5.4\% for the test case described in Sec.~\ref{sec:test}. %

In some cases, the MPI-process communication may govern computational time consumption and the scaling limit. Reducing the number of MPI processes per node, while keeping a constant problem size, increases the domain size per MPI process while reducing the MPI communication and memory usage for the halo data employed for the boundary conditions. Using fewer MPI processes per node leads to a reduction of used resources, if only MPI is employed for the parallelization. Implementing a hybrid parallelisation approach that combines MPI with OpenMP allows a reduction of MPI processes while still utilizing all available cores. FS3D was previously parallelized with OpenMP to run on vector computers and will run on GPU machines in the near future. We have since enhanced and revised the OpenMP parallelisation in FS3D's most time-consuming routines of the evaporation solver by implementing OpenMP at the loop level. As a result, multiple costly MPI communications are no longer necessary, and the shared memory among OpenMP threads reduces the amount of MPI halo data blocking memory. This at first suggests the usage of only one MPI process per ccNUMA node. %
However, other hardware limitations must be considered. For example, on HPE Apollo (Hawk), four CPU cores (CCX) share a common Level 3 (L3) cache, but beyond that caches are not shared. As tested and concluded in a previous report \cite{stober2023dns}, four OpenMP threads per MPI process are considered optimal on the chosen hardware. Eight OpenMP threads do not share the Level 3 cache, resulting in costly memory transfers due to the absence of a "first-touch" principle implementation. Consequently, the following performance study utilizes four OpenMP threads per MPI process.
\vspace{-4mm}

\subsection{Test-case description}\label{sec:test}

\vspace{-4mm}

\begin{figure}[ht]
\centering
\includegraphics[width=0.55\linewidth]{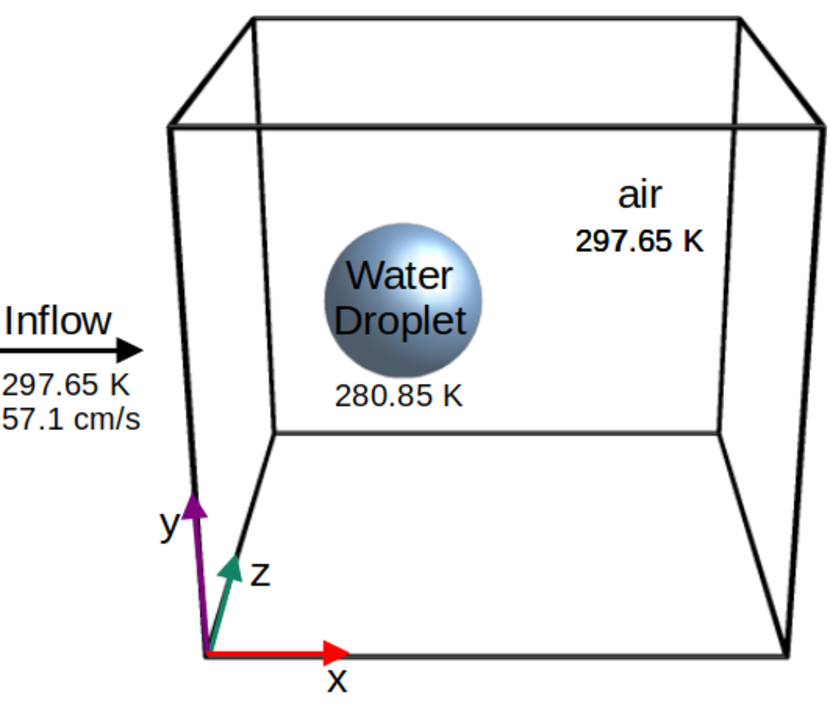}
\caption{Computational domain of the test case used for performance evaluation of the evaporation routines in FS3D.}
\label{evap_domain}
\vspace{-8mm}
\end{figure}

Evaporation of a liquid droplet in a gaseous cross-flow is considered for the performance test of evaporation (or liquid-gas phase change) routines.
The computational domain of the same is shown in Figure \ref{evap_domain}. 
A spherical droplet with an initial temperature of $280.85$~K, lower than the ambient air, is placed in a cubic domain containing the continuous phase (air). 
The initial ambient temperature of the air is $297.65$ K. The droplet is also subjected to a cross flow of the air having a velocity of $0.571$~m/s and a temperature of $297.65$~K, which is the same as the initial temperature of the ambient air. 
Hence, on the left face of the computational domain, the inflow boundary condition and on the right face, a pressure outlet boundary condition is applied. 
The other faces of the domain are subjected to continuous far-field boundary conditions.

\vspace{-4mm}

\vspace{-4mm}

\subsection{Computational time consumption evaluation}
\vspace{-8mm}

\begin{figure}[ht]
\centering
\includegraphics[width=0.67\linewidth]{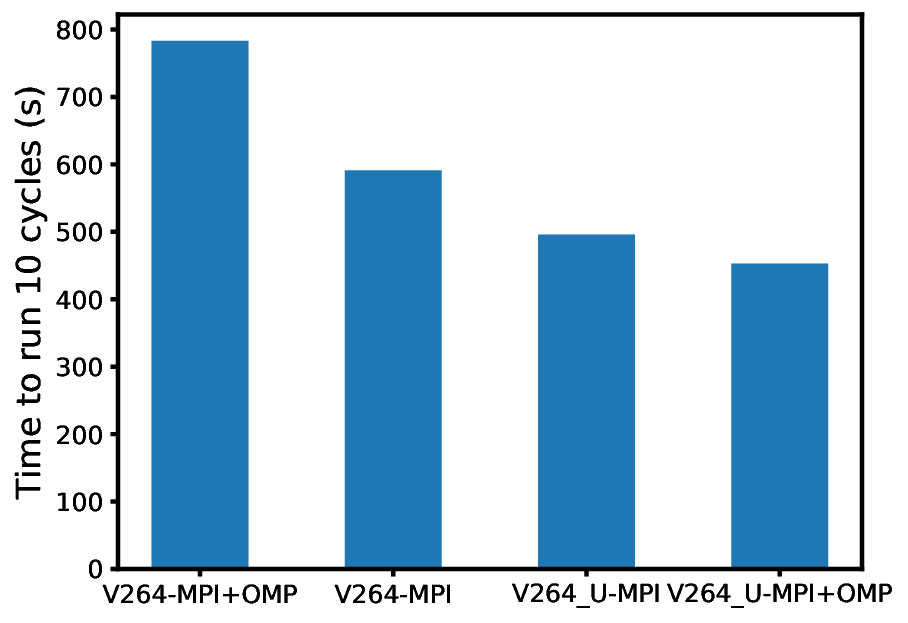}
\caption{Performance of different version of FS3D tested for running 10 time cycles.}
\label{perf}
\vspace{-4mm}
\end{figure}

The performance enhancements were implemented in FS3D version V264. 
The updated version is named V264\_U in this report for convenience. 
Both versions were compiled with the GNU compiler (version 10.2.0) using MPI with OpenMP (MPI+OMP) and without OpenMP to compare the performance in each case.
The performance analysis tool MAQAO \cite{maqao} was used to evaluate the performance of the different versions of FS3D on HPE Apollo (Hawk). 
The size of the grid for the further analyzed test case was 256$\times$256$\times$256. 
128 MPI processes in the case of only MPI and 32 MPI processes with 4 OpenMP threads in the case of MPI+OMP were employed for the analysis with MAQAO on a single node of HPE Apollo (Hawk). 
This ensured the testing of both versions was performed on the same computational resources. 
Figure \ref{perf} shows the total time taken to compute 10-time cycles with the different versions of FS3D. 
It can be observed from the bar chart in Figure \ref{perf} that version V264\_U takes less computational time compared to V264 in both cases with and without OpenMP. 
Also, the MPI+OMP parallelized version of V264\_U results in the lowest observed time to solution. %
This shows, that the introduced optimizations discussed above are beneficial to the overall compute time of FS3D in this specific case. %
The following investigates the scaling behavior of the different versions presented above. %

\vspace{-4mm}

\subsection{Strong scaling performance}

\vspace{-4mm}

\begin{figure}[ht]

\begin{subfigure}{0.49\textwidth}
\includegraphics[width=\linewidth]{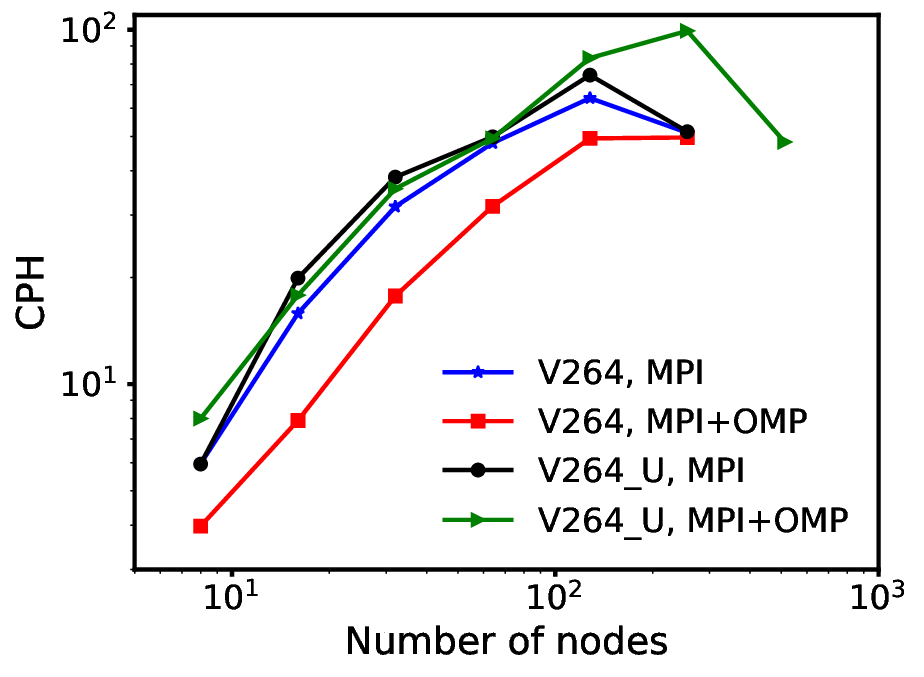}
\caption{}\label{ssplot}
\end{subfigure}
\hspace*{\fill}
\begin{subfigure}{0.49\textwidth}
\includegraphics[width=\linewidth]{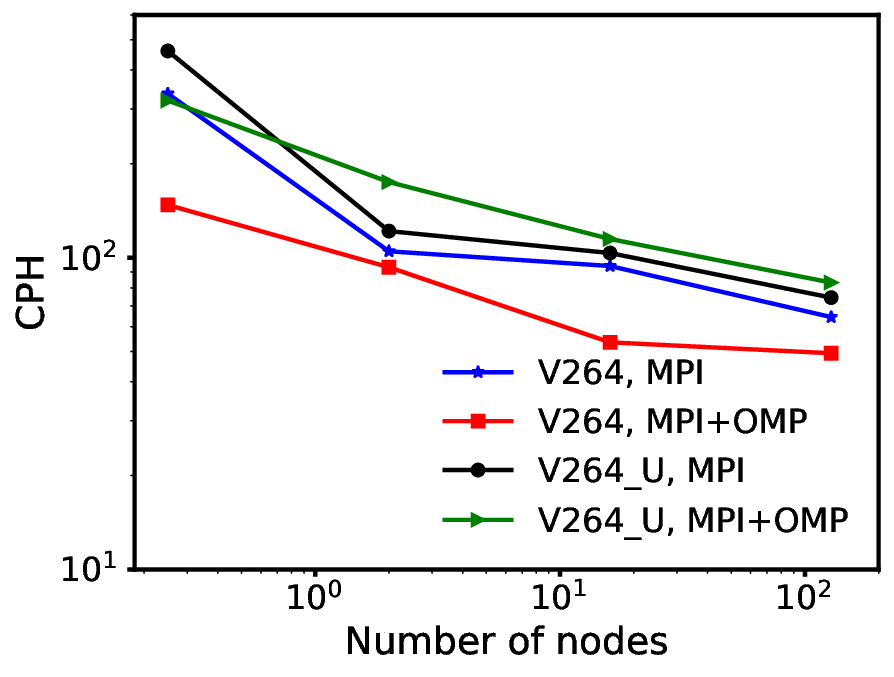}
\caption{}\label{wsplot}
\end{subfigure}

\caption{(a) Strong scaling performance, (b) Weak scaling performance: Estimated cycles per hour (CPH) in case of utilizing MPI and MPI with OpenMP for both versions V264 and V264\_U.}

\vspace{-4mm}
\end{figure}

The effect of increasing the number of cores on the performance is measured by performing a strong scaling, i.e., keeping the grid size fixed while increasing the parallelization. 
A fixed spatial resolution of 1024$\times$1024$\times$1024 grid cells was chosen for the following. %
Figure \ref{ssplot} shows the measured strong scaling performance on a log-plot. 
Figure \ref{ssplot} shows cycles per hour (CPH) calculated by FS3D for the same test case as described in section \ref{sec:test} using two modes of parallelization, i.e., MPI and MPI+OMP. 
The CPH represents the time steps (cycles) simulated per hour. 
The test cases' simulations were made to run for 30 minutes to measure and extrapolate CPH. 
Note that each node of HPE Apollo (Hawk) has 128 CPU cores. 
In the MPI cases without OpenMP, the simulations were performed with one MPI process per core using full nodes. 
For cases using MPI with OpenMP, a parallelization with four OpenMP threads per MPI-process and 32 MPI processes per node was employed, hence again the full nodes were employed. 
Each MPI-process is pinned with a stride of four for the four OpenMP threads. This arrangement allows for a comparison of the usage of the same computational resources, solving the same problem size with an identical resolution, but using different parallelization techniques and FS3D versions with and without the newly introduced optimizations. 
The number of nodes for parallelization was systematically increased to measure the CPH until the CPH started to decline for increasing allocated resources for each combination of the FS3D version and the parallelization technique. 
The largest number of nodes employed for the strong scaling investigation was 512 nodes, i.e., 65536 CPU cores.

% \begin{figure}[ht]
% \centering
% \includegraphics[width=0.75\linewidth]{figures/strong_plot.png}
% \caption{Strong scaling performance: Estimated cycles per hour (CPH) in case of utilizing MPI and MPI with OpenMP for both versions V264 and V264\_U.}
% \label{ssplot}
% \end{figure}

Figure \ref{ssplot} shows that the CPH increases monotonically for all combinations until 128 cores. 
V264 with MPI+OMP shows the lowest CPH compared to other combinations. 
V264\_U with only MPI shows the best performance with 16 and 32 nodes, while V264\_U with MPI+OMP shows the best performance for all other numbers of nodes. 
All combinations except V264\_U with MPI+OMP start to show a performance decrease with 256 nodes. For V264\_U with MPI+OMP, the CPH increases until 256 nodes and then decreases with 512 nodes. 
This shows that the hybrid implementation not only improves the performance for previously used numbers of nodes but also improves the strong scaling limit of usable resources by a factor of two. The maximum CPH attained with V264\_U is 35\% higher compared to V264. %

\vspace{-4mm}

\subsection{Weak scaling performance}
\vspace{-4mm}

In the weak scaling, the number of cells per processing core is kept constant. 
Hence, with an increase in the number of processing cores (or nodes), the number of cells was increased with the number of CPU cores employed. 
The same test case for strong scaling was used to perform a weak scaling test in the present work using 65536 cells per CPU core. 
Each MPI+OMP parallelization had four times the number of grid cells in MPI parallelization so that each core had the same number of cells to process.
Thereby, the same computational resources were used to process the same problem size for all combinations. 

% \begin{figure}[ht]
% \centering
% \includegraphics[width=0.75\linewidth]{figures/weak_plot.png}
% \caption{Weak scaling performance: Estimated cycles per hour (CPH) in case of utilizing MPI and MPI with OpenMP for both versions V264 and V264\_U.}
% \label{wsplot}
% \end{figure}

Figure \ref{wsplot} shows the weak scaling. 
It can be observed that the updated version V264\_U with MPI+OMP computes more CPH than the original version V264 for both types of parallelization techniques when running with 4 to 128 nodes. In the case of 4 nodes, the CPH of V264\_U with MPI+OMP was 67\% higher than V264 with MPI.
However, in the case of a quarter node (32 CPU cores), V264\_U with MPI performs the best, indicating better computation efficiency by a pure MPI parallelization, if not the full node is employed and communication is not the bottleneck anymore.

\vspace{-4mm}
\section{Conclusions}
\vspace{-4mm}

The wetting dynamics of droplets impacting onto a heated surface are presented using DNS simulations. 
The simulations were performed using the in-house DNS code FS3D. 
FS3D can be used for high spatial and temporal resolution as it has been parallelized using MPI and OpenMP. 
A grid-independence study was conducted to ensure that the results were not influenced by the grid resolution.
A grid size of 512$\times$512$\times$512 was found to be sufficient enough to adequately resolve the impact dynamics of droplets considered in the present study. 
The impact dynamics of droplets were investigated for the range of Weber numbers from 27 to 250 and Reynolds numbers from 2230 to 6705. 
For this range of We and Re, the droplet wets the surface smoothly in a spreading regime without showing any splashing. 
As the droplet spreads, it forms a rim at the edge of the droplet as it attains the maximum diameter. 
The maximum wetted diameter of the droplet increases linearly with Re$^{1/5}$.
It is also observed that during spreading, only a very thin portion of the liquid film is heated, indicating that the heat conduction timescale is slower than the inertial timescale of the process for the parameter variation studied here.

In some cases, droplet impact on the heated surface may involve strong evaporation dynamics with a smaller time scale of evaporation. 
Therefore, it is necessary to increase the overall performance of the code for longer simulation times.
The "MAQAO" tool was used on HPE Apollo (HAWK) to measure the computational time required by a new optimized version (V264\_U) and the original version (V264) of FS3D to run 10 time-cycles with two parallelization techniques, i.e., MPI and MPI+OMP. 
The V264\_U version with MPI+OMP shows the best performance with a decrease of 23\% in computational time. 
Two optimizations were introduced: The hybrid parallelization using MPI with four OpenMP threads per MPI process, reduces the overall MPI communication time and memory usage for halo data. This accounts for a reduction of 17.6\%. 
Employing a re-organization of a loop using a red-black scheme leads to a 5.4\% reduction of FS3D's overall runtime. 
Strong and weak scaling of both versions were also evaluated. The V264\_U version showed improved scaling by a factor of 2 compared to V264. 
The V264\_U version can scale efficiently to 256 nodes with no drop in the performance.

%%%%%%%%%%%%%%%%%%%%%%%%%%%%%%%%%%%%%%%%%%%%%%%%%%%%%%%%%%%%%%%%%%%%%%%%%%%%%

\newenvironment{acknowledgments}%
{\null\begin{center}%
 	\bfseries Acknowledgments\end{center}}%
{\null}
\begin{acknowledgments}
The authors kindly acknowledge the {\it High Performance Computing Center Stuttgart} (HLRS) for support and supply of computational resources on the HPE Apollo (Hawk) platform under the Grant No. {FS3D/11142}. Additionally, the authors gratefully acknowledge the financial support of the Deutsche Forschungsgemeinschaft (DFG, German Research Foundation) under Germany's Excellence Strategy EXC2075-390740016 through the project 409029509, "Investigation of Droplet Motion and Grouping" and through the project 270852890, GRK2160: "Droplet Interaction Technologies (DROPIT)".\\
\end{acknowledgments}

\vspace{-8mm}

\bibliography{FS3D_HLRS2024}
\bibliographystyle{spmpsci}
%\bibliographystyle{spbasic}
%\bibliography{references}

%\input{references}

\eject
\end{document}